\documentclass{aanda}
\usepackage{graphicx}

\usepackage{txfonts}
\usepackage{xcolor}
\usepackage[normalem]{ulem}
\usepackage[breaklinks,colorlinks=true,linkcolor=blue,citecolor=blue]{hyperref}
\begin{document}

   \title{Bursty acceleration and 3D trajectories of electrons in a solar flare }
          
   \authorrunning{Bhunia et al.}
   \author{Shilpi Bhunia
          \inst{1, 2}, Laura A. Hayes\inst{3}
          ,
          Karl Ludwig Klein\inst{4,5}
          ,
          Nicole Vilmer\inst{4,5}
          , Shane A. Maloney\inst{1}
          \and Peter T. Gallagher\inst{1}
          }

   \institute{Astronomy \& Astrophysics Section, Dublin Institute for Advanced Studies, Dublin, D02 XF86, Ireland.
         \and
             School of Physics, Trinity College Dublin, College Green, Dublin 2, Ireland.\\ 
            \email{bhunias@tcd.ie}
           \and
          European Space Agency, ESTEC, Keplerlaan 1 - 2201 AZ, Noordwijk, The Netherlands.
          \and
         LIRA, Observatoire de Paris, PSL Research University, CNRS, Sorbonne Université, Université Paris Cité, 5 place Jules Janssen, 92195
Meudon, France
          \and
          Observatoire Radioastronomique de Nançay, Observatoire de Paris, PSL Research University, CNRS, Univ. Orléans, 18330 Nançay, France
             }

   \date{Received ; accepted }

 
  \abstract
   {During a solar flare, electrons are accelerated to non-thermal energies as a result of magnetic reconnection. These electrons then propagate upwards and downwards from the energy release site along magnetic field lines and produce radio and X-ray emission.}
   {
    On 11 November 2022, an M5.1 solar flare was observed by the Spectrometer/Telescope for Imaging X-rays (STIX) on board Solar Orbiter together with various ground- and space-based radio instruments. The flare was associated with several fine hard X-ray (HXR) structures and a complex set of metric radio bursts (type III, J, and narrowband). By studying the evolution of X-ray, extreme ultraviolet, and radio sources, we aim to study the trajectories of the flare-accelerated electrons in the lower solar atmosphere and low corona.
    }
   {We used observations from the STIX on board Solar Orbiter to study the evolution of X-ray sources. Using radio imaging from the Nançay  Radio heliograph (NRH) and the Newkirk density model, we constructed 3D trajectories of 14 radio bursts. }
   {Imaging of the HXR fine structures shows several sources at different times. The STIX and NRH imaging shows correlated changes in the location of the HXR and radio source at the highest frequency during the most intense impulsive period. Imaging and 3D trajectories of all the bursts show that electrons are getting accelerated at different locations and along several distinct field lines. Some of the trajectories from the same origin show expansion on the order of 4 over a height of $\sim$110 Mm. The longitude and latitude extent of the trajectories are $\sim30\arcsec$ and $\sim152\arcsec$.}
   {We find that the electrons producing HXR and radio emission have similar acceleration origins. Importantly, our study supports the scenario that the flare acceleration process is temporally and spatially fragmentary, and during each of these small-scale processes, the electron beams are injected into a very fibrous environment and produce complex HXR and radio emission.}


   \maketitle
%

\section{Introduction}
During solar flares, magnetic reconnection in the solar atmosphere can release up to $10^{32}$ ergs of free magnetic energy in a matter of minutes \citep{2012Emslie}. Most of this energy is transferred to non-thermal electrons that then propagate away from the acceleration site along the magnetic field lines, both upwards away from the Sun and downwards towards the chromosphere. 
According to the standard flare model, following acceleration, the population of electrons travelling downwards along magnetic field lines hit the chromosphere, losing their energy via Coulomb collision and giving rise to hard X-ray (HXR) footpoints through non-thermal bremsstrahlung. This process heats the plasma at the flare foot points, causing the upflow motion of the plasma along the flare loops and resulting in soft X-ray (SXR) emission \citep{2006Milligan, 2006Milligan2}. Through observations of the temporal, spectral, and spatial profiles of the X-ray photons from space-based observatories, information about these accelerated energetic electrons can be inferred. Hence, there is much ongoing research on the spatial and temporal characteristics of the acceleration origin of solar flares.
It is suggested that the particle acceleration takes place in the corona and sometimes HXR sources are observed around the acceleration site \citep{2010krucker}, but often they are too faint to observe in the presence of the bright footpoint sources with current instrumentation.

Flare-accelerated electrons also propagate anti-sunwards away from the acceleration site and propagate upwards along open magnetic field lines and produce type III radio bursts \citep{Reid_review2014}. Sometimes electron beams can also propagate along closed field lines and produce type J and U \citep{Reid_2017_J}. Among these bursts, type IIIs are commonly observed in the frequency range from a few 100 MHz to 10 MHz and can be observed in the interplanetary medium down to 100 kHz or even lower, called an `interplanetary' (IP) burst.
These radio bursts are generated when electron beams generate Langmuir waves leading to radio emission at the fundamental ($f_p$) and second harmonic ($2f_p$) of the plasma frequency, given by $f_p \approx 8980\sqrt{n_{e}}$ Hz, where $n_e$ is electron number density (cm$^{-3}$). 
As the plasma density decreases with increasing coronal height, the propagation of accelerated electron beams away from the Sun results in the drift of emission from high to low frequencies.
The fast drift of these bursts is thought to be due to the rapid propagation of energetic electron beams \citep{1950Wild}. 
Using imaging spectroscopy we can examine the trajectory of these radio bursts and hence track the electron beams from very low in the corona to the upper corona.
Many studies \citep[][]{2008Klein, Chen_2013,2018Mann,2018Chen,2019Zhang} have imaged type III bursts at different frequencies to track the propagation of the electron beams.

Sometimes HXR emissions have been observed simultaneously with type III bursts at decimetric and metric wavelengths (refer to \cite{Pick2008} for a review) which has suggested that these two kinds of beam populations generating HXR and radio emission have a common accelerating origin. 
This temporal association between these two types of emissions has been studied both for individual events \citep[e.g.][]{1982Kane, 1995Aschwanden, Vilmer2002} and multiple events \citep[e.g.][]{Kane1972,1981Kane, 1995aAschwanden, Arzner2005, 2017Reid, 2023James}. A statistical study of the simultaneous HXR and type III bursts associated with coronal jets was performed by \cite{Krucker_2011}. It showed that the interchange reconnection model (reconnection between closed and open field lines) is the most common magnetic topology leading to these events. However past theoretical works \citep{1984Sturrock,1992Benz,1998Isliker,2012Cargill} proposed that the acceleration processes in solar flares are much more complex and result from a collection of multiple fragmented energy releases. Sometimes, the HXR flux of the flares is observed with many peaks and pulsations, and followed by multiple fine structures (including spikes, type IIIs, and Js) in radio emission at decimetric and metric wavelengths \citep{2011Reid,2014Reid}, which are possibly observational evidence of this theory.
Previous modelling work done by \cite{1995Vlahos} on multiple beam propagation based on the assumption of fragmented acceleration processes and fibrous corona found various fine structures of type IIIs and decimetric spikes. A previous study by \cite{Chen_2013} investigated type III bursts in decimetric wavelengths using high-resolution imaging-spectroscopy of the Karl G. Jansky Very Large Array (VLA). The authors tracked the sources of the bursts in the gigahertz (GHz) range; that is, very close to the reconnection site. They found distinct multiple beam trajectories within $\lesssim$ 1 s revealing the fibrous magnetic structures of the solar corona and further supporting this idea of the acceleration process being highly fragmentary temporally and spatially. Another study of type IIIs in decimetric wavelengths by \cite{2018Chen} using the same instruments shows tens of fast electron beams travelling along multiple field lines with different topologies and dynamical properties (such as the density profiles along the field line) within the timescale of 50~ms. Also, these electron beams came from a very compact region (<600~km$^{2}$), which again supports a fibrous reconnection region. It is important to realise that a type III burst at MHz frequencies does not come from an individual electron beam travelling along a single well-defined magnetic field line, but from a bundle of field lines and therefore from a probably large volume within the heliosphere \citep{1989Roelof_pick}. This is important when analysing large radio sources in the context of investigating the propagation of electron beams at these heights.

To understand flare particle acceleration process(es), and to relate the acceleration of electrons both away from and towards the Sun, insights can be gained by using combined HXR and metric/decametric radio observations that have high temporal and spatial resolution. Investigating radio bursts with HXR emission allows us to track the flare-accelerated electrons from the chromosphere to very high in the corona and it is particularly interesting to compare the spatial observations of HXR emission with radio bursts \citep{Vilmer2002} which can provide vital importance to our understanding of the acceleration process(es) that occur in solar flares. The Spectrometer/Telescope for Imaging X-rays \citep[STIX;][]{2020krucker} on board Solar Orbiter \citep{2020Muller} has an excellent time resolution of up to 0.5 s and can provide images with an angular resolution down to 7\arcsec over a 2° field of view in the 4--150~keV energy band, making it an ideal instrument with which to study solar flare time variability.
Often, individual interplanetary type III bursts are found to be associated with multiple type IIIs in the low corona \citep{2005Klein,2017Reid,2023James}, suggesting the propagation of multiple electron beams into the interplanetary medium. Hence, we need high temporal radio imaging-spectroscopy in metric wavelengths to investigate the propagation of multiple electron beams.

In this paper, we investigate groups of bursts of radio and HXR emission associated with a M5.1 GOES-class solar flare that occurred on 11 November 2022 and that was observed by multiple instruments including STIX, the Atmospheric Imaging Assembly  \citep[AIA;][]{2012Lemen} on board the Solar Dynamics Observatory \citep[SDO;][]{2012Pesnell}, and ground-based radio instruments such as the Irish Low Frequency Array \citep[I-LOFAR;][]{2021Murphy}, Observations Radio pour Fedome et l’Etude des Eruptions Solaires \citep[ORFEES;][]{orfees} radio spectrograph, and Nançay Radioheliograph \citep[NRH;][]{1997Kerdraon}. By combining extreme ultraviolet (EUV), X-ray, and radio observations we investigate the emissions associated with the acceleration process(es) and track the accelerated electron beams in the low solar corona. In Section~\ref{sec:overview} we present the event overview. In Section~\ref{imaging} we present the combined observations from EUV/UV and X-ray to investigate the UV ribbons and the X-ray sources. We show radio sources of different bursts overlaid on AIA images and in 3D to understand the topology of the trajectories. We discuss the observations in the context of reconnection and particle acceleration in Section~\ref{discuss} and conclude in Section~\ref{conclude}.

\section{Event overview} \label{sec:overview}
\begin{figure}[ht!]
   \centering
      \includegraphics[width = 8.5cm]{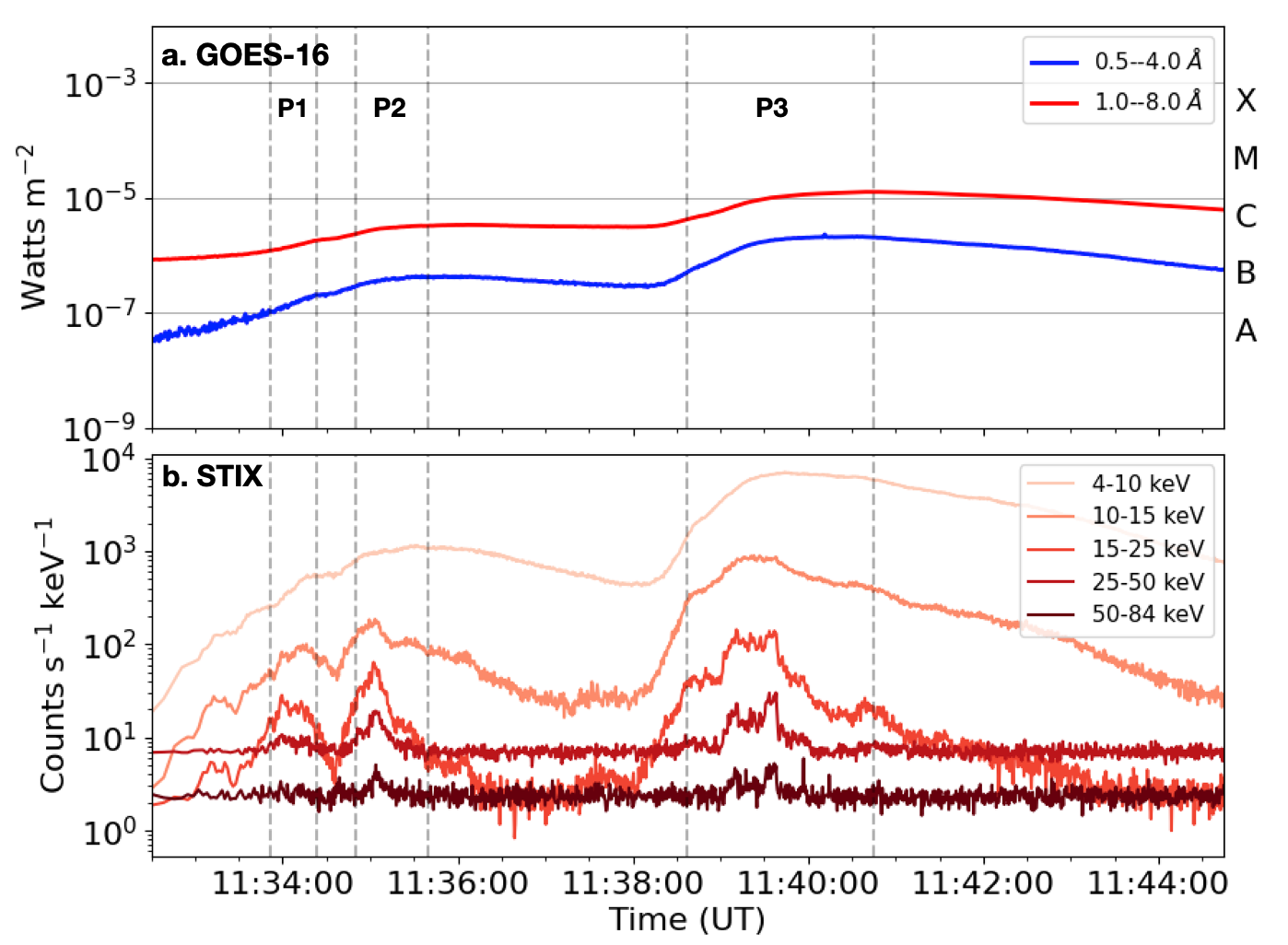}
      \caption{(a)  GOES-16 light curves in two pass bands (1-8~\AA\ and 0-5~\AA) indicating an M class flare on 11 November 2022. (b) STIX light curves in five different energy bands show the same flare. The three major non-thermal periods (P1, P2, and P3), indicated by the dotted lines, show multiple fine structures.} 
         \label{fig:goes_stix}
        \end{figure}

\begin{figure*}[ht!]
   \centering
      \includegraphics[width = 15cm]{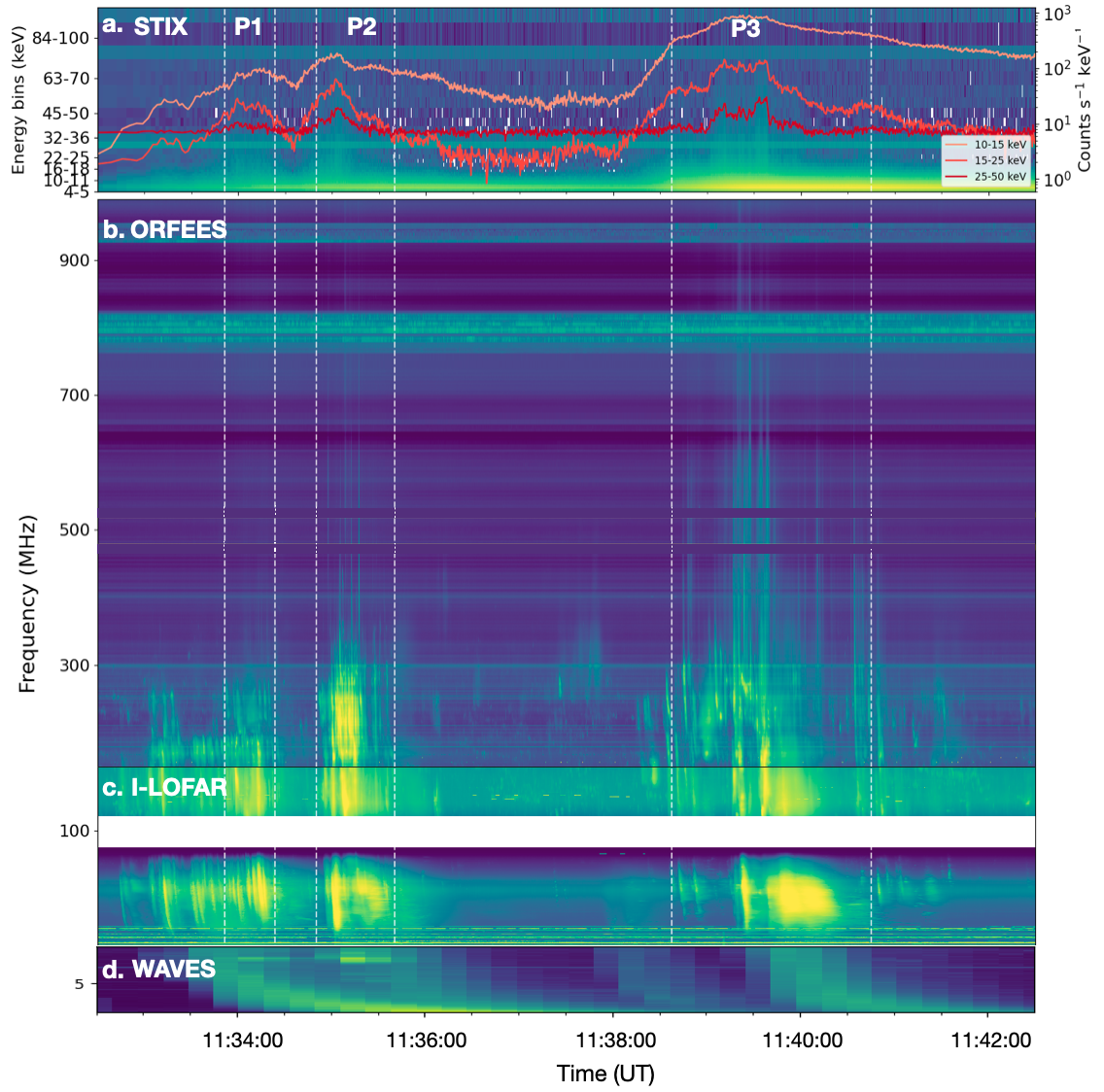}
      \caption{(a) STIX X-ray spectrogram with the time series of the summed non-thermal energy bands overplotted on top of it. The plot shows the three major non-thermal periods.
      (b), (c), and (d) show the type III bursts observed by ORFEES, I-LOFAR, and WAVES in the frequency range from 1 - 1004~MHz. Clearly there is a temporal association between each individual HXR period and groups of radio bursts.}
         \label{fig:radio_spectra}
        \end{figure*}

On 11 November 2022, the X-ray Sensor (XRS) on board the Geostationary Operational Environmental Satellite-16 (GOES-16) detected an M1.2 class solar flare (Fig~\ref{fig:goes_stix} (a)) that started at 11:32:36\,UT from the active region (AR) NOAA 13141 located at 211\arcsec, 199\arcsec (Fig~\ref{fig:flare_loc} a) described in Section~\ref{sec:tracking_bursts}). Following the flare, no CME was observed in LASCO C2/C3 suggesting that there was no large-scale eruption. However, there were lots of plasma outflows associated with the flare, as can be observed in the AIA EUV images (the bottom panel of Fig~\ref{fig:pfss_brsts}).
Solar Orbiter was at a heliocentric distance of $\sim0.62$ AU and 21.8 deg to the east from the Sun-Earth line. Hence for this event, STIX had a shorter photon arrival time compared to near-Earth and ground-based instruments. The STIX time series (Fig~\ref{fig:goes_stix} (b)) has been corrected to the light travel time of Earth.
Fig~\ref{fig:goes_stix} (b) shows STIX light curves of summed counts for five different summed energy bands. There are clearly three major non-thermal periods (P1, P2, and P3) seen in the HXR energy ranges (15-25~keV, 25-50~keV, and 50-84~keV). The time range of each period is indicated by the vertical dotted white lines. There are also several HXR spikes seen within each period, especially during P3.
Fig~\ref{fig:radio_spectra}a shows the STIX X-ray spectrogram, with the time series overplotted in red, again showing the three non-thermal periods.
        
In addition to multiple HXR flux peaks, the flare also produced many type III bursts in the GHz to MHz range observed by multiple ground-based radio instruments such as I-LOFAR, ORFEES, and space-based instrument the Radio and Plasma Wave Investigation \citep[WAVES;][]{1995waves} on board the WIND spacecraft. Fig~\ref{fig:radio_spectra} b, c, and d show the combined radio spectra from ORFEES, I-LOFAR, and WIND/WAVES in the frequency range from 1 - 1004 MHz. There were no observations available in the high GHz range. 
Both I-LOFAR and ORFEES spectra show many different spectral features of coronal type III bursts and some type J bursts (which will be discussed in Section~\ref{P2} and Section~\ref{P3}).  The WAVES spectra show that these type IIIs are observed as low as $\sim$1~MHz, indicating the propagation of multiple electron beams in the interplanetary medium.
Comparing the X-ray spectrogram and time series with radio spectra from ORFEES and I-LOFAR, one can see that there exists a good temporal association between the individual HXR period at 15-25~keV and the groups of radio bursts. Even outside of these periods, it is seen that whenever the HXR flux increases there are radio bursts associated with it.
This motivates our study of the physical relationship between these two kind of emissions. 
In general, it is seen that during each non-thermal period as the HXR emission increases the starting frequency of the associated radio bursts also increases.
During P1, the highest starting frequency for the radio emission was $\sim$289~MHz. During P2 and P3, the highest starting frequency is 770.3 and 920.42~MHz, respectively. This is because the increase in the HXR flux implies that there are more high energetic electron beams, and hence for these beams it is much easier and faster to produce Langmuir waves, leading to the generation of radio emission at much lower heights; that is, at higher starting frequencies. This is most likely why we also see more bursts that are brighter at higher frequencies (>500~MHz) during P3 compared to the ones during P2.

\section{Observations and data analysis}\label{imaging} 
\subsection{Complex ribbon formation and X-ray sources}
\begin{figure*}[ht!]
   \centering
      \includegraphics[width = 13cm]{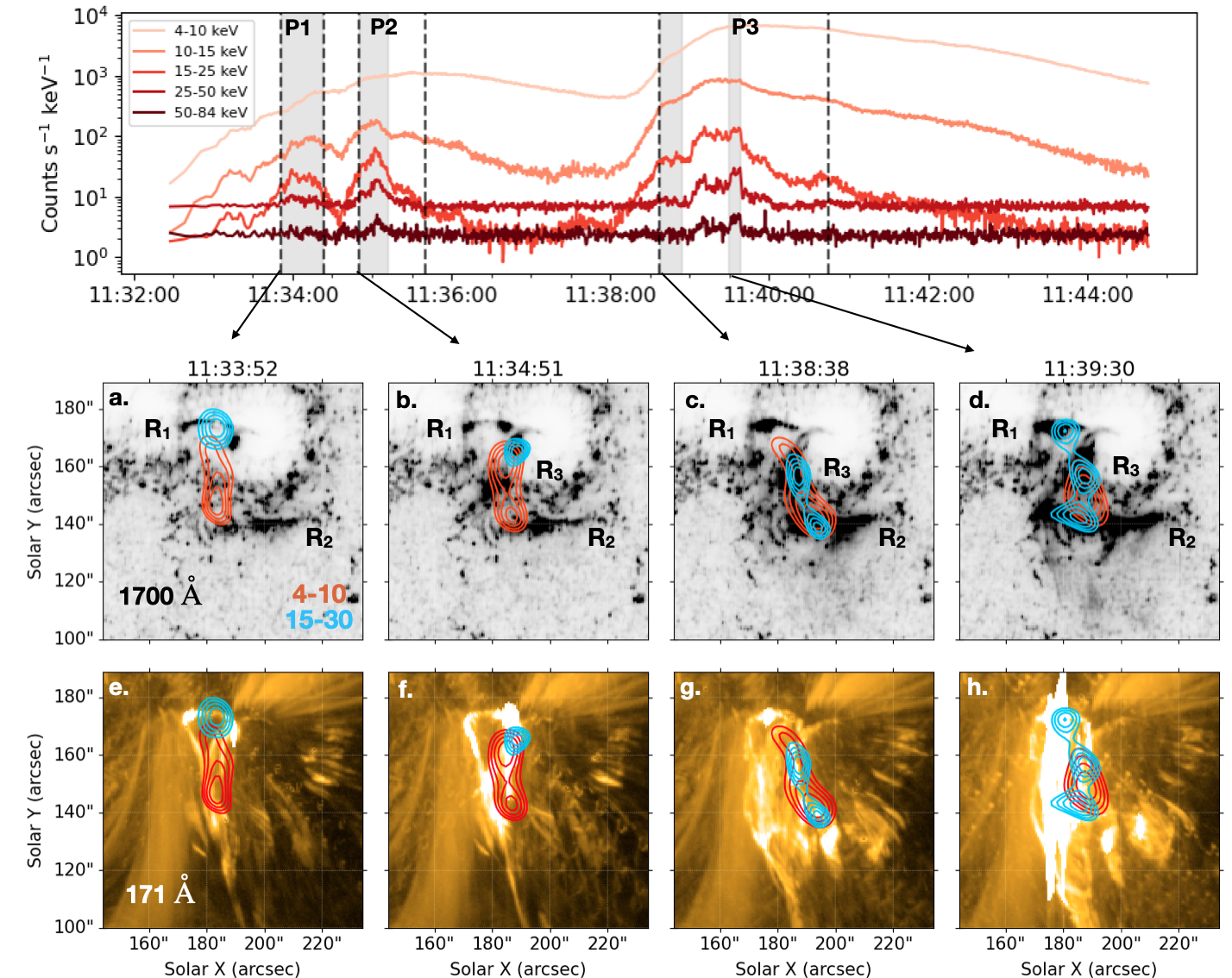}
      \caption{The top panel shows the STIX timeseries in five energy bands. The middle and bottom panels show the flare evolution in 1700~\AA\ and 171~\AA, respectively, during each of the non-thermal peaks. Overlaid on top of each image are the reprojected thermal (4-10 keV, red) and non-thermal (15-30 keV, sky blue) contours of the STIX images using the MEM\_GE method. Panels a, b, c, and d show the appearance of one, two, or three non-thermal sources at different locations during this complex ribbon formation whereas panels e, f, g, and h show the hot loops and plasma outflows.
       } 
         \label{fig:euv}
        \end{figure*}

To understand the flare in the lower atmosphere, we investigated the EUV/UV images from AIA and X-ray observations from STIX.
The top panel of Fig~\ref{fig:euv} shows the STIX time series in five energy bands. The middle and bottom panels of Fig~\ref{fig:euv} show the evolution of the flare in the 1700~\AA\, and 171~\AA\ AIA images, respectively. To investigate each of the major HXR non-thermal peaks, image reconstruction for the STIX pixel science data was performed using the IDL-based STIX ground analysis software \citep{ewan_2024}. The MEM\_GE imaging algorithm \citep{Massa_2020} was used to construct images of the X-ray sources during these peaks. 
The start of the integration time used to construct the images is shown in the title of the 1700~\AA\ ~images. The STIX thermal (4-10 keV, red contours) and non-thermal (15-30 keV, sky blue contours) sources have been reprojected to the Earth observer viewpoint and plotted on the EUV images. The integration times for STIX thermal and non-thermal contours for each panel of Fig~\ref{fig:euv} are 32, 22, 17, and 9 s and are indicated as the grey-shaded region on the time series. The integration times are different to allow for better counting statistics during each period to produce the images.
The contour levels are 60, 70, 80, and 90\% of the peak intensity. 

During the first non-thermal peak at 11:33:52\,UT,  two ribbons are observed, labelled as R1 and R2, in the 1700~\AA\ image (Fig~\ref{fig:euv} (a)). One non-thermal source appears on R1 and in the 171~\AA\ ~image  (Fig~\ref{fig:euv} (e)), we see a few brightenings on R1 and a loop-like structure. The thermal source is most likely positioned on the loop top. At the same time interval, some narrowband drifting radio emissions and many type IIIs are observed indicating several acceleration episodes (see Figure~\ref{fig:radio_spectra}). 

At the second non-thermal peak at 11:34:51 UT, one can see the appearance of the third ribbon (R3) in between R1 and R2 (Fig~\ref{fig:euv} (b) and (f)). The non-thermal source moves from the previous position to the top right corner of R3 which is also evident from the brightening at the same place in 171~\AA ~image. Possibly, we have a second set of loops connecting R2 and R3. The thermal contours possibly show the loop legs from the top.

At the start of the third non-thermal peak at 11:38:38 UT,  two non-thermal X-ray sources are observed located on top of R3 and R2. It looks like the direction of both HXR and SXR sources is towards the southwest.
The observations at 171~\AA\ at the same time show the start of the large-scale plasma motion.  At 11:39:30 UT, the 1700~\AA\ image (Fig~\ref{fig:euv} (d)) shows the appearance of three non-thermal sources appearing on each of the three ribbons whereas the 171~\AA\ image shows intense brightening, indicating strong heating and energy deposition is happening at each of these ribbons. Also in the 1700~\AA\ image, the R3 is observed to extend towards the east; simultaneously, the HXR source on top of it also extends. Hence, the morphology of the magnetic structure has likely changed based on the location of the ribbons and HXR sources.

In summary, during this flare, multiple acceleration episodes are evident which are indicated by the non-thermal peaks. X-ray imaging during these non-thermal peaks shows one, two, or three HXR sources at different locations implying that there are several injections of electrons at different locations. This further indicates that during each of these periods, magnetic topology may have changed.

\subsection{Tracking beam trajectory} \label{sec:tracking_bursts}
To track the accelerated electrons in the upper corona that produce the type III emission, and compare their location with the HXR emissions in the chromosphere to that in the high corona, the locations of the radio bursts were imaged. We used NRH to image the bursts as I-LOFAR and ORFEES do not have the imaging capabilities.
Next, the detailed spectra were examined to identify several radio bursts to track the sources in frequency as they propagate through the corona. 
To achieve this we chose bursts for which we could image at least five of all the available NRH frequencies (432, 408, 327.5, 327, 298.7, 270.6, 228, 173.2, and 150.9~MHz). The processing of NRH data was undertaken by using Solar SoftWare (SSW) \citep{1998Freeland} NRH software at a time resolution of 0.25~s. We generated 256 $\times$ 256 pixels (each pixel size $\approx 15.17\arcsec$) of 2D intensity images for each frequency within its field of view$\approx$$2^{\circ}$$\times$$2^{\circ}$. The main uncertainty of the imaging observations at metre wavelengths comes from variations in the refractive index in the ionosphere, due to electron density variations that are induced by gravity waves in the underlying neutral atmosphere\citep{Mrc-86,Mrc-96}.
\begin{figure*}[ht!]
   \centering
      \includegraphics[width = 12cm]{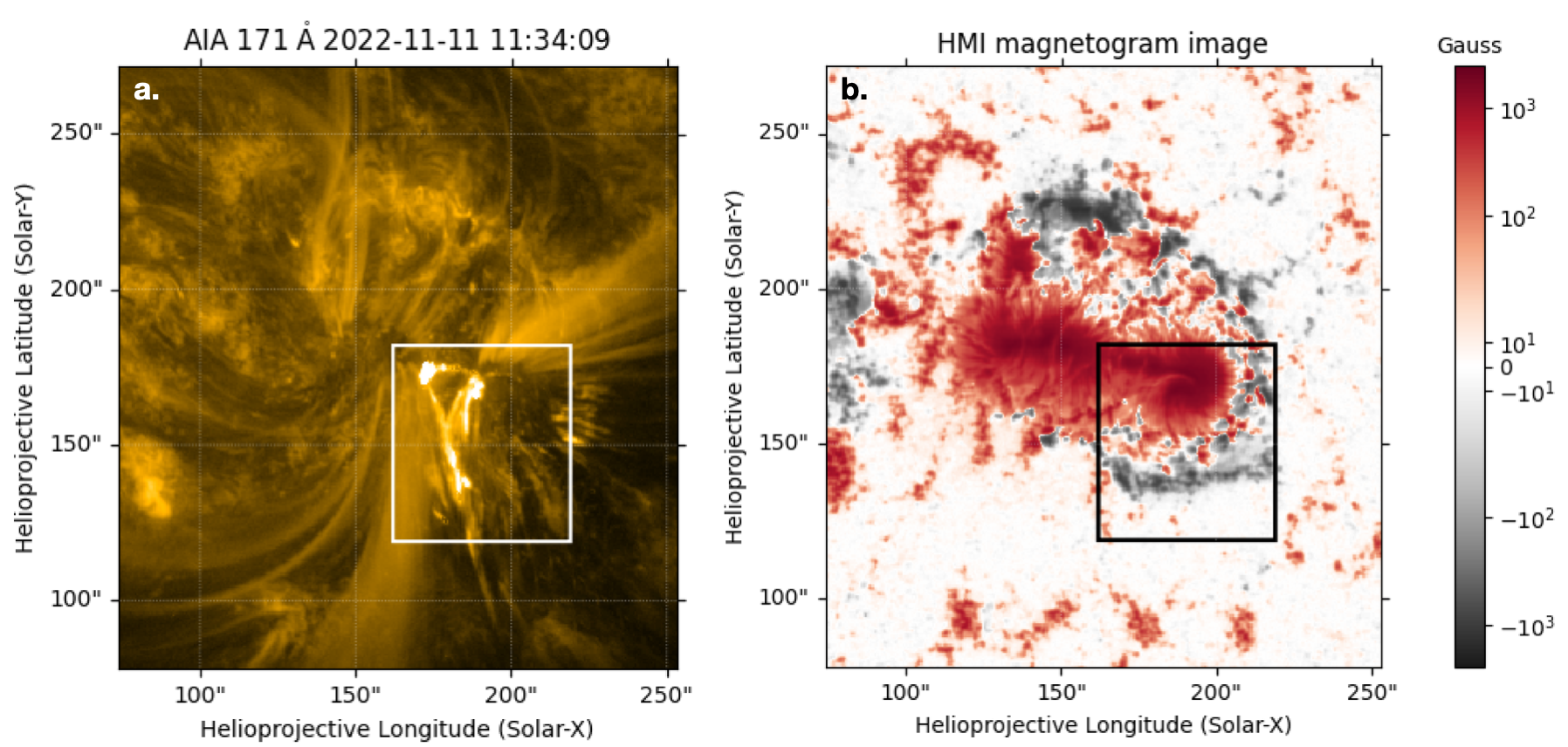}
      \caption{The left panel (a) shows the AIA 171\AA~image at 11:34:09 UT. The flare region is indicated by the white square box. The right panel (b) shows the HMI magnetogram image of the flare region. The flaring region has lots of emerging negative field lines.} 
         \label{fig:flare_loc}
        \end{figure*} 
\begin{figure*}[ht!]
   \centering
      \includegraphics[width = 17cm]{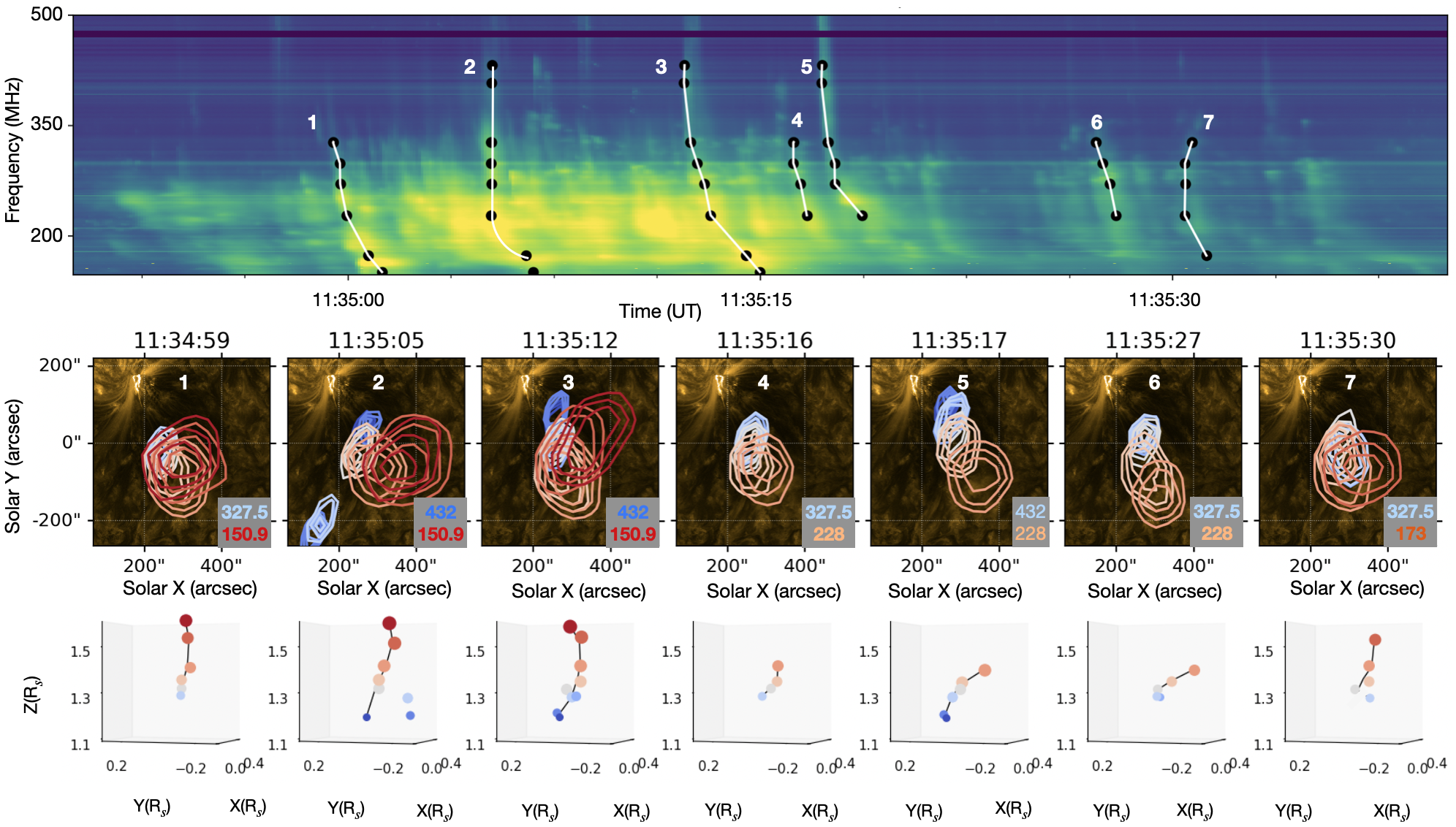}
      \caption{Top panel: the dynamic spectra from 11:34:50 - 11:35:40 UT during P2 (2nd HXR non-thermal period). The identified bursts for imaging are indicated by the white lines on the spectra. The black dots are the frequencies and times at which NRH imaged the bursts. We identify two type Js (B2 and B5), type IIIs (B1 and B3), some narrowband emissions (B4 and B6), and one bidirectional drift burst (B7).
      Middle panel: shows the radio contours (85, 90, and 95 \% of the peak intensity) of each burst overlaid on the AIA 171\AA\ image. Bottom panel: the position of the centroids of each burst in 3D. Both the middle and bottom panels show that even if the source location at each frequency is different, in general the trajectories have curvature.  } 
         \label{fig:radioperiod2}
        \end{figure*}
The variations make the apparent source position undergo variations on timescales of several tens of minutes. To quantify these variations in source locations that are due to the ionospheric variations, and to estimate the errors in the source positions we analysed a noise storm on 11 November 2022, which is non-thermal emission from the parent active region that lasts for at least the seven-hour observing time of the NRH. The extremes of the deviation in the apparent noise storm location around the type III bursts of interest were observed around 11:23 and 11:49 UT. We fitted the measurements of the centroid positions of the noise storm in this time interval, excluding the range 11:30-11:45 UT in which the type III bursts occurred, to a sine function and derived the ionospheric deviations at the times of the type III bursts from the fit. During the time intervals of the type III bursts the deviations induced by the ionosphere at 150.9 MHz range from -0.03 to +0.08 R$_\odot$ in the east-west direction and from  -0.02 to +0.10 R$_\odot$ in the south-north direction. The uncertainty of the correction from the statistical uncertainty of the fit to the sine function is $\pm 0.11$ R$_\odot$. The effect decreases with the square of the frequency ratio at higher frequencies. We therefore did not correct the position of the type III bursts but will discuss its influence in the description of imaging observations.
        
We omitted P1 for this analysis as during this period we see some narrowband drifting emissions starting around 280 MHz and ending at ~215 MHz and using NRH we can only image them at 228 and 270.6 MHz, which is not enough to track the radio sources.
We hence only focussed on choosing bursts during the P2 and P3 time periods. To locate the type III bursts on the open field lines we first investigated the magnetic topology near the flaring region we looked at the HMI magnetogram image of the flaring region (Fig~\ref{fig:flare_loc} b)). The image shows that the flare happened at the edge of the active region (indicated by the box) surrounded by many opposite polarity emerging field lines. Then 
using the pfsspy software \citep{Stansby2020}, the Potential Field Source Surface (PFSS) extrapolation was performed to locate the open magnetic field lines (top panel of Fig~\ref{fig:pfss_brsts}).
 The PFSS extrapolation shows no clear open field lines near the active region of interest. However, there must be open lines as the type III bursts were also observed in the WIND/WAVES spectra (Fig 2 (d)). We conclude that the current-free condition of the PFSS model is not satisfied in this very active region, where eight flares, including one of class M, had occurred before the type III bursts of interest. It does, however, show that closed field lines (gray) from the flaring region are connected to a large null point far away (large loops).
To better understand the trajectories of these bursts, we attempted to estimate the positions of the radio sources in 3D. To achieve this, first, we assumed that all the target bursts are harmonic, as these bursts start above 327\,MHz and the observation of fundamental emission around this frequency is rare due to free-free absorption \citep{2000Robinson}. Also, the circular polarisation of type III bursts is usually low at NRH frequencies \citep{Mrc-90}, although this does not seem to always be the case \citep{2020Rahman}. We cannot use this constraint, because NRH does not measure the circular polarisation in its present version (since 2020). 
To calculate the radial distance of the sources ($R$), we used the four-fold Newkirk density model \citep{1961Newkirk} for very dense coronal active regions. Then, using the on-disc positions ($x,y$) of the radio sources, we could estimate the z positions of the sources by using the equation, $z = \sqrt{R^{2} - r^{2}}$, where $r = \sqrt{x^{2} + y^{2}}$
. After calculating the 3D positions of NRH source centroids, we then connected the centroids by a spline-fitting curve representing the trajectory of the type III sources. We consider this curve to be the average of the field lines along which the electron beams propagate and produce these type III sources. The main uncertainty comes from the ionospheric density variations, as was described before. They will affect the lower frequencies (228 - 150.9 MHz), but only on timescales of several minutes. Using a different density model and fold number for the model will result in different $z$ heights of the sources at a given frequency. However, irrespective of which fold we use for the Newkirk model, the curvature of the line is not affected. This is because the curvature of the line is majorly influenced by the on-disc co-ordinates of the sources.

\subsubsection{Radio bursts during P2}\label{P2}

The top panel of Fig ~\ref{fig:radioperiod2} shows the detailed spectra of radio bursts from 500 to 147~MHz during P2 (11:34:50 - 11:35:40 UT). One can see lots of bright features among which we have picked the bursts 
based on the variations in drifting behaviour, bandwidth, and starting frequencies and can image at at least five of all the NRH frequencies. On the spectra, the chosen bursts are marked by white lines connecting the NRH frequencies but the bursts can start at higher frequencies.
The black dots are the preferred times at which the bursts are considered to be best imaged at NRH frequencies. The middle panel of Fig ~\ref{fig:radioperiod2} shows NRH radio contours (85, 90, and 95 \% of the peak intensity) of each burst overlaid on AIA images. The numbers on each image indicate the burst number written on the spectra. The start and end frequencies at which the bursts were possible to image are indicated in the figures of the middle panel. The bottom panel shows the 3D trajectory of the radio sources of the same bursts.
One can see that the bursts have different morphologies, brightnesses, and drift rates. The starting frequencies vary from 770.3 to 327.5~MHz. While the details of the source morphology may be distorted by the ionospheric refraction at frequencies below about 250 MHz, the changes within timescales of a few seconds are not.

Some of these bursts (labelled `B` and burst numbers B2, B3, and B5) have very high drift rates at high frequencies (>350 MHz). B2 starts at 633~MHz and ends at 150.9 MHz. We see a very fast drift at first and 
the burst appears to have a bright extended emission from 145 MHz to 110 MHz, appearing like a J burst. From the image and the 3D trajectory, one can see that there is a bending of the position of the radio sources at 173 and 150.9~MHz. This statement is independent of ionospheric refraction, which does not exceed $20\arcsec$ at 150.9 MHz in the time range of bursts B1 to B7. Plotting the radio sources on top of the PFSS field lines (Fig~\ref{fig:pfss_brsts} a.) also shows the trajectory of the sources to be closely lying on top of the large loops. To examine the relationship between the location of the plasma outflow and the burst sources, we overplotted radio sources of this burst on the running ratio image of the flare produced by AIA 171, 193, and 211\AA (Fig~\ref{fig:pfss_brsts} e.). It is seen that the radio source at 432~MHz is located in the direction of the plasma outflow.
Similarly, B5 also appears to be a J burst. It starts at 759~MHz and appears to have a very fast drift. Around 250~MHz, it has a bright feature, giving it the appearance of a 'hockey stick'. The image and the 3D trajectory show that there is a curve in the trajectory. B3 appears to be a type III burst, starting at ~730 MHz. Similarly, like previous bursts it tends to have a fast drift rate, and then from around 300~MHz the drift starts to slow down. 

Interestingly, many bursts start to appear brighter within the frequency range 298-357 MHz. This can happen if the electron beams have a similar energy distribution and have to travel similar distances
before producing the bump in tail instability leading to the generation of intense radio emission at similar frequencies.
Among these bursts, there are many narrowband bursts (B4, B6) and some type IIIs (B1) that drift to low frequencies. B1 starts at 405\~MHz and extends to 27 MHz (seen in the I-LOFAR spectra). On the dynamic spectra of this burst, one can also see fragmented bright structures from 277 to 150~MHz. The starting frequencies of B4 and B6 are 330 and 352 MHz, respectively. 
 Interestingly, even though the source locations are not the same, the trajectories look similar for B1, B4, and B6 for the 432-228~MHz frequency range. 
 B7 looks like a bidirectional drift burst given that one can see a positive frequency drift from 298 to 360 MHz. Usually, this kind of feature is the signature of electrons going downwards in the corona. Unfortunately, with NRH we can only image at 327.5 and 327~MHz; hence, we cannot infer much about the trajectories of the electron beam in this frequency range.
At a given frequency, the location of the bursts is not at the same place. For example, at 435 MHz the location of B2, B3, and B5 is (258\arcsec, 30\arcsec), (242\arcsec, 45\arcsec), and (258\arcsec, 91\arcsec), respectively. At this high frequency, the ionospheric variations do not affect the source locations. The differences in the locations of bursts that follow each other in close succession show that the electron beams producing successive bursts travel along different closed and open field lines.
Comparing the 2D images and the 3D trajectories of these bursts, it is evident that all of these trajectories curve towards the west. The source position of the highest frequency for all the bursts appears to the south-west of the flaring region. It appears that the location of the HXR source on ribbon R2 and the jet influence the starting location of the burst in the south-west direction from the flaring region (look at panel b of Fig~\ref{fig:euv} and panel e of Fig~\ref{fig:pfss_brsts} together). 

\subsubsection{Radio bursts during P3}\label{P3}
\begin{figure*}[ht!]
   \centering
      \includegraphics[width = 17cm]{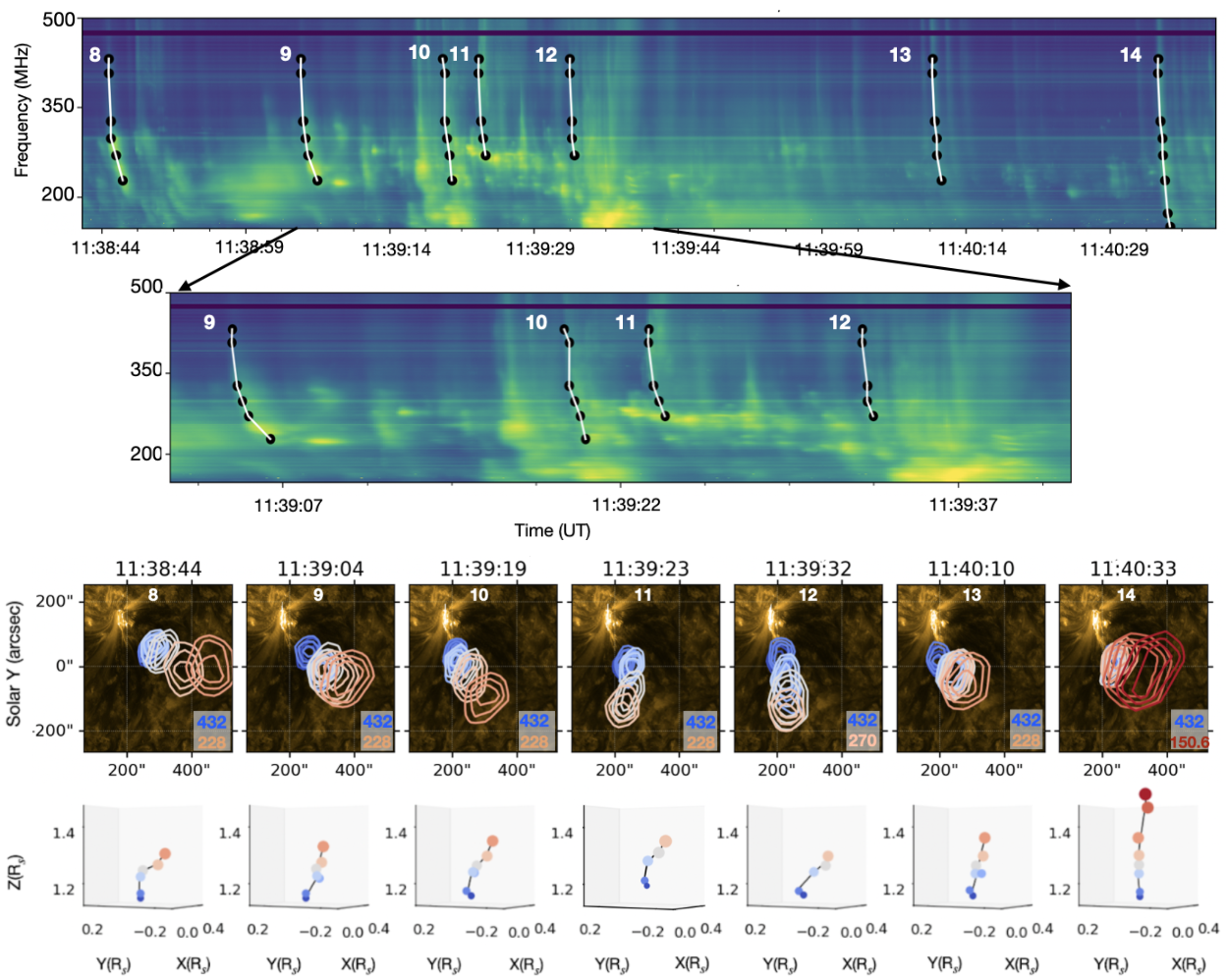}
      \caption{The top panel shows the dynamic spectra from 11:38:42-11:40:38 UT. We also show the zoom-in of the spectra showing that the bursts are different than the ones during P2. Here we identify three different types of bursts: a) the type IIIs with sharp cutoff (B8 \& B9), b) diffuse type IIIs (B10, B11 \& B12) \& individual type IIIs (B13 \& B14). The third panel from the top shows the sources for each burst overplotted on AIA images. It clearly shows that the trajectory of each burst is quite different. The trajectories in 3D show that they are in general different looking than the ones in P2.  } 
         \label{fig:radioperiod3}
        \end{figure*}

The top panel of Fig ~\ref{fig:radioperiod3} shows the detailed spectra of radio bursts from 500 to 147~MHz during P3 (11:38:42 - 11:40:38 UT). We identified three different kinds of type III bursts depending on the spectral features: bursts with a sharp cut-off (B8), type J burst (B9), clusters of type IIIs (B10, B11 and B12), and individual type IIIs (B13 and B14). These bursts look very different than the identified bursts during P2, indicated by the zoomed-in spectra of 40~s (same time width as the P2 spectra). 

B8 starts at 560~MHz and has a sharp low-frequency cut-off at 205~MHz. This can happen when the electron beam loses its energy very fast \citep{1995Vlahos}. For the B9 burst, the cut-off is at 228 MHz and the drift tends towards 0, implying it is possibly a type J burst. Fig~\ref{fig:pfss_brsts} b and f show the radio sources to be lying on top of the large loops and in the direction of the plasma outflow.
The imaging shows that for these two bursts, radio sources at 432~MHz  originate from a similar location south-west of the flaring region. 


Next, from 11:39:16 UT to 11:39:40 UT there are clusters of bursts and during this period the HXR flux is also observed to be most intense, implying that there are more energetic electron beams available.
Among the bursts, we have identified B10, B11, and B12 in the regions of clusters of type IIIs. The time of the imaging for these bursts was
chosen by tracking the leading edge of the spectral feature of these bursts. B10 has a very high starting frequency of 920~MHz (see Fig~\ref{fig:radio_spectra} b). The trajectories of each burst appear different. Also, from this period onwards, the sources at 432~MHz appear to be located more southwards relative to the flaring region than the previous bursts. 
During this time, a similar change in the location of the HXR source on ribbon R2 has also been observed (compare Fig~\ref{fig:euv} c and d). One can also see a similar shift in the direction of the plasma outflow (compare Fig~\ref{fig:pfss_brsts} f and g).

From 11:39:58 to 11:40:38 we find multiple individual type IIIs. Among them, we picked B13 and B14. B13 has a starting frequency of 645 and ends at 228~MHz. B14 has a high starting frequency at 842 MHz and a frequency drift of 275~MHz/s (within the frequency range of 842 - 111~MHz). It looks like the beams were mostly moving toward the observer. PFSS field lines (Fig~\ref{fig:euv} d) show the burst sources to be lying on the closed field lines.
In general, all of these bursts have less sharp curves than some of the bursts during P2.

\subsubsection{3D trajectories}
To examine the trajectories more closely, we have plotted the trajectories of the bursts during P2 and P3 in Fig~\ref{fig:field_lines} a, and b, respectively. Different propagation paths are indicated by different colors and circles on individual trajectories marking the source locations at different frequencies for the bursts. The blue circles on the XY plane mark the position of HXR sources during each period. Both figures show clear evidence that the trajectories of the exciters differ between individual bursts. This indicates that after acceleration electron beams travel along different closed or open field lines.
Fig~\ref{fig:all_brsts} displays the 3D trajectories of all 14 bursts under two different viewing angles. The right plot shows the clear positional shift of the radio sources at 432~MHz of bursts (B10, B11, B12, B13, and B14) during P3 indicated by the black arrow. At 432 MHz, the average separation between these bursts and B9 is $\sim 49 \arcsec$. The ionospheric variations do not cause such a large separation at this high frequency. As was mentioned in the previous section, a positional shift is also seen in the HXR source overlaid on R2 (compare Fig~\ref{fig:euv} c and d).
Therefore the general difference in the source location of these bursts compared to the previous bursts suggests that the trajectory of the electron beams getting accelerated and producing these successive type III bursts has shifted.

\section{Discussion}\label{discuss}
\subsection{Summary of the observation results} \label{summary} 
\begin{figure*}[ht!]
   \centering
      \includegraphics[width = 14cm]{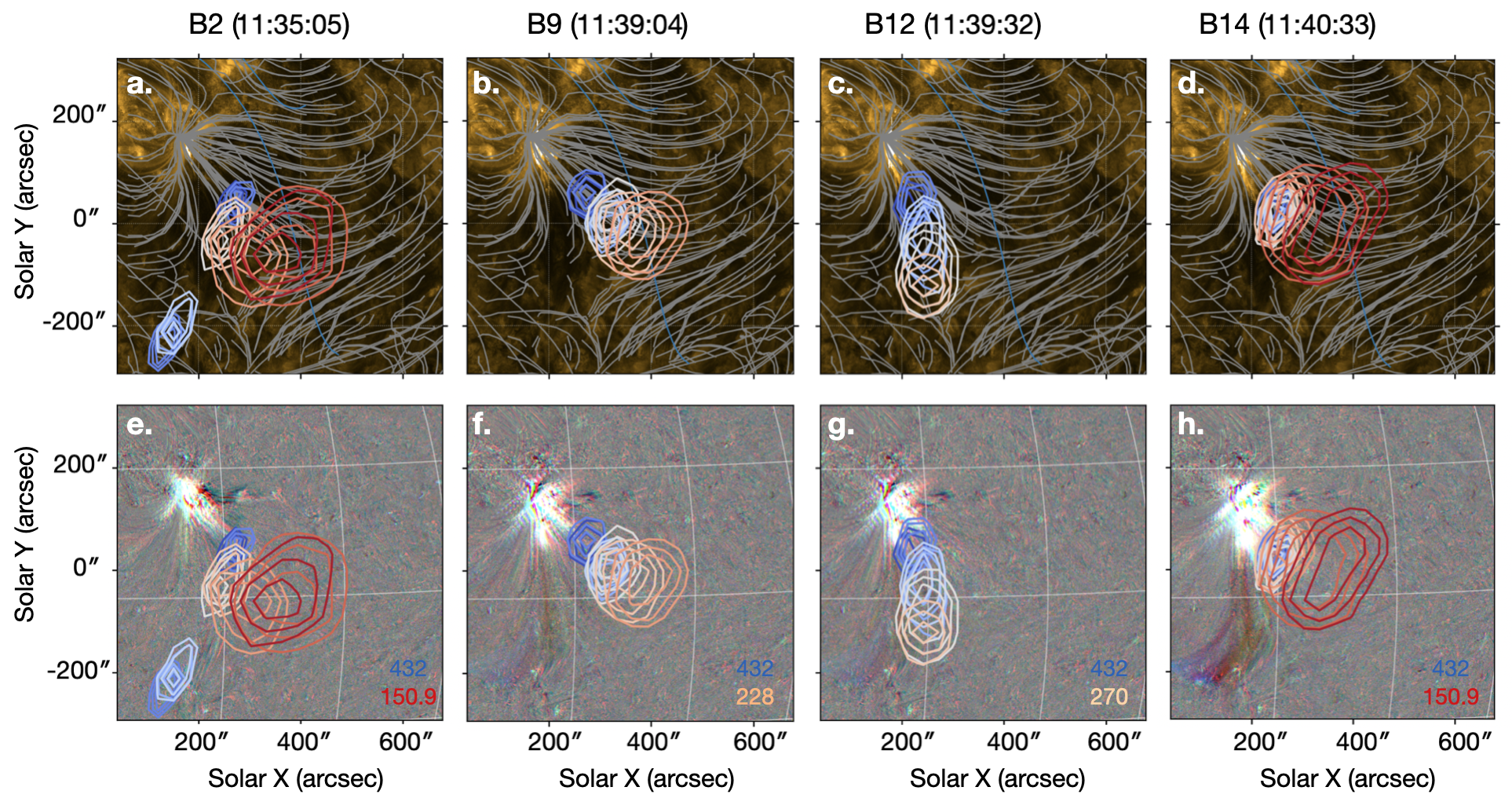}
      \caption{PFSS field lines are overlaid on the AIA 171\AA ~image with B2, B9, B12, and B14 at different times (indicated on the title)
      The extrapolation was done using the Global Oscillation Network Group (GONG) magnetogram synoptic map observations at 09:44 UT. In the bottom panel, we have overlaid the same radio bursts on the AIA ratio images of 211\AA, 193\AA, and 171\AA ~to look at the plasma outflows. } 
         \label{fig:pfss_brsts}
        \end{figure*}

\begin{figure*}[ht!]
   \centering
      \includegraphics[width = 14cm]{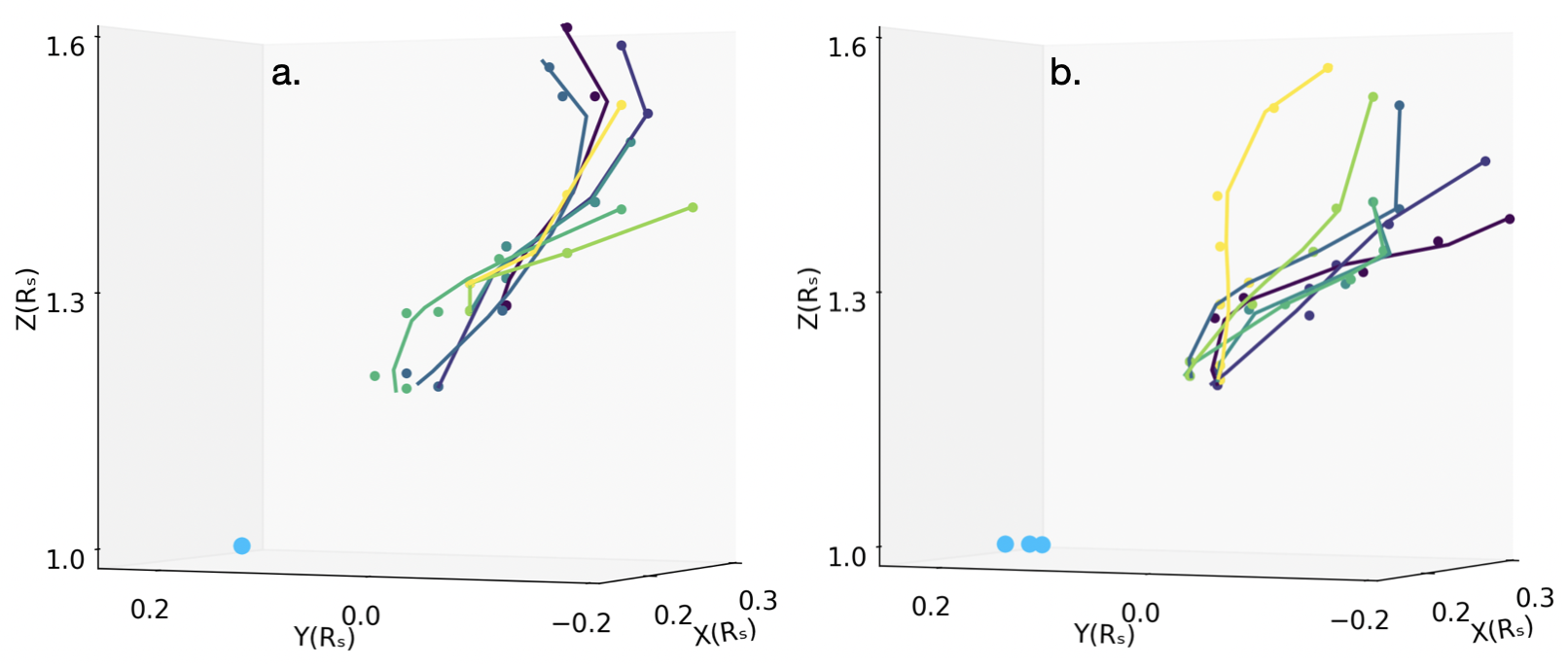}
      \caption{The left (a) and right panel (b) show the trajectories of the radio bursts during P2 and P3, respectively. Different colour implies different bursts and the circles imply the location of the sources. The blue circles imply the locations of the  HXR sources on the XY plane, respectively. One can clearly see the topology of the trajectories is quite different between P2 and P3.} 
         \label{fig:field_lines}
        \end{figure*}

\begin{figure*}[ht!]
   \centering
      \includegraphics[width = 14cm]{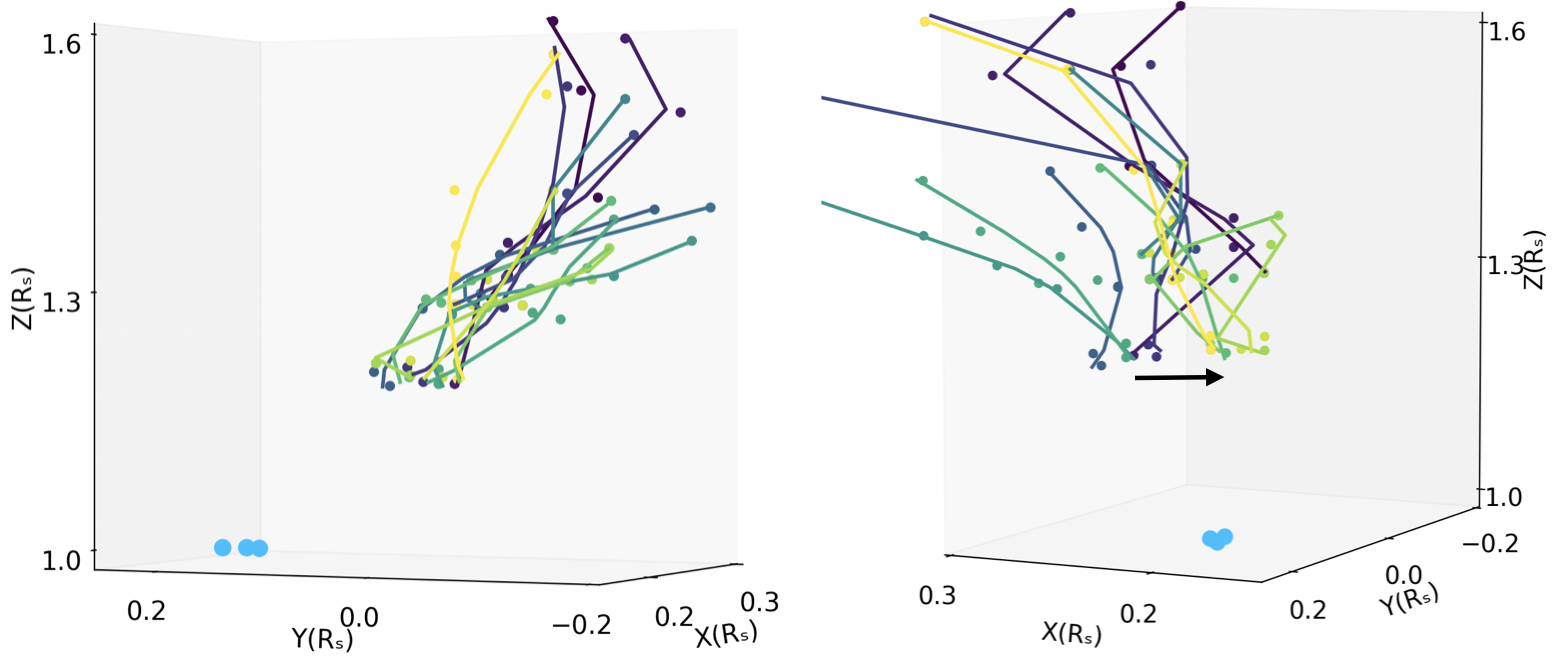}
      \caption{The left and right plots show the 3D trajectories of all 14 radio bursts under two different viewing angles. In the right figure, one can see the shift of the radio sources at 432 MHz for bursts during P3 indicated by the arrow.} 
         \label{fig:all_brsts}
        \end{figure*}

In this paper, we have investigated the observational signatures of accelerated electrons during an M1.2 GOES class solar flare on 11 November 2022 using Solar Orbiter/STIX, AIA/SDO, and multiple ground- and space-based radio instruments. Here are the key findings:
\begin{enumerate}
   
    \item The flare produced HXR emission with
    multiple bursts and fine structures (Fig~\ref{fig:radio_spectra}(a)). There are three major non-thermal periods. During these periods, multiple radio bursts (type IIIs, Js, and narrowband) with different drift rates, and starting frequencies were observed. Some of the type III bursts were seen extended to 1 MHz (Fig~\ref{fig:radio_spectra}(d)), indicating the escape of flare-accelerated electrons along open field lines into the interplanetary medium.
    \item  During each period as the HXR intensity increases, the starting frequency of the radio bursts also increases (Figure~\ref{fig:radio_spectra}). During P1, P2, and P3, the highest frequency of the radio bursts is found to be 289, 770.3, and 920.42 MHz, respectively.
    \item STIX and AIA imaging (Fig~\ref{fig:euv}) show multiple HXR sources on three ribbons indicating several electron beam injection episodes at different locations. 
    \item During the most intense non-thermal period (P3), the radio source at 432~MHz shifts from south-west to the south of the flaring region (compare the location of the sources of bursts B9 and B10 in Fig~\ref{fig:radioperiod3}). A similar change can be seen in the location of the ribbon (R2) and the HXR source on top of it (Fig~\ref{fig:euv}(c) and (d)). This indicates the change in the acceleration site and supports the idea that the HXR and radio-emitting electrons have a common acceleration origin.
    
    \item During each of these non-thermal periods the AIA running ratio images (bottom panel of Fig~\ref{fig:pfss_brsts}) show plasma outflows away from the flare site and the sources at 432 MHz are located near the trajectory of the plasma outflows.
    
    \item We tracked the sources of 14 bursts. During each period, the 2D location of the radio sources is different at the same frequencies indicating that the accelerated electron beams producing the radio busts had access to different bundles of field lines. 
    \item Among the bursts, radio bursts with different morphology are observed. 
    If the electron beams are confined to large closed field loops they produce type J bursts (B2, B5 \& B8). On some open field lines, the beams propagating escape from the corona and contribute to type III bursts (B1, B14). On other field lines, the beams stop, possibly generating type III-shaped spectra over a limited frequency band (B8, B13) or narrowband features (B4, B6).  

    \item While variations in the ionospheric refractive index do affect imaging observations at lower NRH frequencies (228, 173.2, 150.9 MHz), we checked that the findings in points 6 and 7 above are not artefacts. This is shown by abrupt changes in the source morphology on timescales (some seconds to a few tens of seconds) that are smaller than those of the ionospheric variations. The bursts also happen to occur at a time when the quasi-sinusoidal variation in the positional shift due to refraction is near zero.

    \item During P2 and P3, the morphology of the 3D trajectories (Fig~\ref{fig:field_lines}) for bursts that follow each other are found to be different implying that the electron beams are travelling along different open or closed field lines. During P3, the location of these trajectories shifts indicating that the location of electrons getting accelerated also changes (Fig~\ref{fig:all_brsts}).
\end{enumerate}
\subsection{Simultaneous observations of HXR emission and radio bursts}
During solar flares, it is often observed that the HXR emissions have bursty features and fine structures, and occasionally appear to have associated radio bursts in decimetric \citep{Benz2005} and metric wavelengths. Often these two kinds of emissions have a good temporal association suggesting that perhaps the acceleration origin of these two kinds of emission is the same \citep[e.g.][]{Kane1972,1982Kane,raoult1985,1995Aschwanden,Vilmer2002, Arzner2005,2011Reid,2017Reid,2023James}. 
As was mentioned, a correlation between HXR intensity and the starting frequency of radio bursts is observed. Two reasons could explain the change in the starting frequencies. One is the change in the height of the acceleration region. The second is a reduced path length for the electrons to develop an unstable beam distribution. Given the correlation with the HXR count rate, the latter explanation is likely to be the cause \citep[e.g.][]{raoult1985,Vilmer2002,2011Reid,2014Reid,2023James}.   
When the HXR intensity increases, 
the number of highly energetic electrons also increases, leading to fast injection
of these electrons. Then, rapid electrons can outpace the slower electrons much faster and produce bump-in-tail instability earlier, leading to radio emission at lower heights \citep{2014Reid_H}. Assuming a static acceleration in the corona, the minimum path length the electrons had to travel from the acceleration height to produce the radio emission can be estimated. From the highest starting frequency (920.42 MHz), the minimum height of the acceleration location was found to be below 1.03 \(\textup{R}_\odot\) when the four-fold Newkirk model was used.

\subsection{Signatures of bursty energy release}
It was suggested in previous literature \citep{Krucker_2011,2012Glesener} that the magnetic topology involving coronal type IIIs and jets and HXR emission mostly involves interchange reconnection whereby the reconnection happens between emerging field loops and open field lines. In this case, three X-ray sources (two at the footpoints of the newly formed closed loop and one at the footpoints of the open field line) are observed.  In our case, we observe several HXR sources (one, two, and three) at different times, and the fact that we also observe type J bursts and multiple narrowband emissions suggests that the acceleration process is far more complex than a simple interchange reconnection model picture. 
It is most likely that the acceleration processes in solar flares result from a collection of many small-scale reconnection events, as was suggested in \cite{1984Sturrock,1992Benz,1995Vlahos,1998Isliker}. Multiple fine structures observed in HXR time series and the spectra of radio bursts are the observational evidence. Moreover, modelling work done by \cite{1989Roelof_pick,1990Pick_vandenoord,1995Vlahos,1998Isliker} shows that the active region is highly inhomogeneous and consists of multiple bundles of field lines. Photospheric motion and emerging flux tubes will lead to lots of fragmented current sheets at separate sites, leading to lots of fragmented energy release processes. This flare occurred at the edge of the active region surrounded by many emerging field lines of opposite polarity. Large coronal loops were observed from the flaring region. Multiple reconnection processes can happen between these overlying large field lines and the newly emerging flux \citep{1999Schrijver}.
This leads to multiple fragmented energy-release processes accelerating non-thermal electrons \citep{1984Sturrock,1992Benz, 1998Isliker,2012Cargill} upwards and downwards from the acceleration sites producing multiplefine, complex structures of radio and HXR emission. 
Several HXR sources indicate multiple electron beam injection episodes. The correlated change in the location of the HXR and radio sources at 432~MHz during P3 suggests that the acceleration origin for two emissions is common and the acceleration regions are multiple. 
 As was mentioned before the radio sources are observed in the direction of the plasma outflows. Similarly to the HXR and 432 MHz source, the outflow becomes intense and shifts southward concerning the flaring site during the third most intense impulsive period. This again supports the idea that electron beam acceleration and plasma outflows occur during multiple reconnection processes between the emerging flux and ambient field lines \citep{1992Shibata,Krucker_2011,2012Glesener, 2016Carley,2017Morosan}

\subsection{Propagation of the electron beams inferred from the radio bursts}\label{bursts}

During each of these small-scale reconnection processes, multiple electron beams with varying energy and number density are injected into various magnetic field line bundles and produce various complex structures of radio emission in the decimetric-metric wavelengths \citep{1995Vlahos}. By tracking different radio bursts (type IIIs, Js, and narrowband), this study shows the beams following distinct field lines with different morphology in 3D, providing observational evidence to the theory that the corona in fibrous \citep{1989Roelof_pick}. The trajectories of all 14 bursts indicate the different acceleration locations of the electron beams.
In the corona, the field lines are expected to radially expand with increasing height \citep{DeForest_2007}. The expansion of the bundle of trajectories originating from similar locations was calculated and found to be expanding from $\sim$ 25 Mm to $\sim$ 111 Mm over a height of $\sim$110 Mm (432 - 270 MHz). 

Often it is seen that multiple coronal type III bursts merge into an individual hectometric-to-kilometric type III burst \citep{1996Poqu,2017Reid}, which is also seen in the Waves spectra. Multiple reasons could be behind this observation. One is that this is perhaps due to the limited time resolution of the instrument. Two, the velocity dispersion of beams can also lead to this merge. However, in a fibrous corona, at a given time type III emission can be generated from a range of locations. 
As a consequence, the emission at a single frequency of a dekametric-to-kilometric type III burst comes from a large volume in space.
To understand the extent to which the electrons were injected into the heliosphere,  we calculated the angular range of the field lines. At 1.37 R\textsubscript{\(\odot\)}, the longitude and latitude of the extent are 30\arcsec $\pm$ 2\arcsec and 152\arcsec $\pm$ 6\arcsec. The angular range is possibly affected by the large beam size of NRH(at 432 and 150 MHz, the major axes are 171\arcsec and 480\arcsec and the minor axes are 78\arcsec and 205\arcsec ).  Hence in the future, we want to use VLA, LOFAR, and MWA, which have excellent spectral and spatial resolution, to track the electron beams and the extension of the field lines. 
By tracking the flare-accelerated electrons from the chromosphere to high in the corona our work adds evidence to support the idea of the corona being very fibrous and the energy release process being highly fragmented in space and time. This idea was previously supported by several publications \citep[e.g.][]{1985Benz,1992Benz,1995Vlahos,1998Isliker,2012Cargill,2020Ramesh}. \cite{Chen_2013,2018Chen} tracked the bursts in the GHz regime and found them to follow very distinct field lines coming from a very compact region. 
 The repetitive plasma outflows suggest the dynamic nature of the acceleration process. In the EUV images, it is not possible to track the outflows due to low spatial and temporal resolution. Future work will be to identify events for which the High-Resolution Imagers (HRI) of the EUI instrument \citep{2020Rochus} on board Solar Orbiter had observations that we can connect with STIX and radio instruments to help study the spatial and temporal variability of the eruption and track the accelerated electrons in the solar atmosphere.

\section{Conclusion}\label{conclude}
In this study, we have tracked the flare-accelerated electrons from the chromosphere to the upper corona using EUV, X-ray, and radio observations. We have found many coronal bursts associated with several HXR non-thermal peaks. Together with EUV and X-ray imaging, we have shown that several HXR sources and repetitive jet-like structures indicate multiple acceleration processes during the flare. 
With the high temporal resolution of ORFEES \& I-LOFAR, radio bursts are observed within the timescale of even <1 s. NRH imaging and 3D trajectories show that electrons are getting accelerated at different locations and propagate along various distinct field lines. Hence, we conclude that the flare acceleration process is very fragmented temporally and spatially. STIX and NRH imaging also show similar changes in the location of HXR and radio sources, indicating that the acceleration source of the X-ray and radio-emitting electrons is common.

\begin{acknowledgements}
Solar Orbiter is a space mission of international collaboration between ESA and NASA, operated by ESA. The STIX instrument
is an international collaboration between Switzerland, Poland, France, Czech Republic, Germany, Austria, Ireland, and Italy. S.B. is supported by the Hamilton PhD Scholarship at DIAS.
L.A.H is supported by an ESA Research Fellowship. The authors would like to acknowledge AIA, GOES, and the WIND team for open access to their data. The authors also thank the Radio Solar Database service at LESIA (Observatoire de Paris) for providing the NRH and ORFEES data.
S.B. acknowledges support from ESA through the ESA Space Science Faculty Visitor scheme - Funding reference ESA-SCI-SA-LE-082. S.B. acknowledges funding support from the French government under the French Research Residency 2023 program. N.V. would like to acknowledge support from CNES. S.M. is supported by an ESA/PRODEX award administered by Enterprise Ireland. This paper made use of an open package called sunpy \citep{sunpy_community2020}.

\end{acknowledgements}

\bibliography{aanda}{}
\bibliographystyle{aanda}

\end{document}